\documentclass[12pt,aps,english]{revtex4}

\usepackage{graphicx}
\usepackage{bm}
\usepackage{dcolumn}
\usepackage[usenames, dvipsnames]{color}
\usepackage{epsfig}
\usepackage{amsmath}

\begin{document}

\title{Chain reconfiguration in active noise}
\author{Nairhita Samanta and Rajarshi Chakrabarti*}
\affiliation{Department of Chemistry, Indian Institute of Technology Bombay, Mumbai, Powai 400076, E-mail: rajarshi@chem.iitb.ac.in}
\date{\today}

\begin{abstract}

In a typical single molecule experiment, dynamics of an unfolded proteins is studied by determining the reconfiguration time using long-range F$\ddot{o}$rster resonance energy transfer where the reconfiguration time is the characteristic decay time of the position correlation between two residues of the protein. In this paper we theoretically calculate the reconfiguration time for a single flexible polymer in presence of active noise. The study suggests that though the MSD grows faster, the chain reconfiguration is always slower in presence of long-lived active noise with exponential temporal correlation. Similar behavior is observed for a worm like semi-flexible chain and a Zimm chain. However it is primarily the characteristic correlation time of the active noise and not the strength that controls the increase in the reconfiguration time. In a nutshell, such active noise makes the polymer to move faster but the correlation loss between the monomers becomes slower.

\end{abstract}

\maketitle

\section{Introduction}

Active processes giving rise to non-equilibrium fluctuations are ubiquitous in biological systems. This is notably distinct from the incessant motion exhibited by particles in any fluid known as Brownian motion which results from the constant collision of the particle with its surrounding solvent molecules \cite{Chandrasekhar}. However in biological systems active motion are driven by the chemical energy produced from the hydrolysis of adenosine triphosphate (ATP). For instance the motion of cytoskeleton inside the cells is controlled by the motor proteins which involves ATP hydrolysis \cite{Weitz}. Other examples would be cell membranes which are perpetually out of equilibrium through active processes \cite{Bassereau} and swimming bacteria which control the active transport of nutrients in aqueous medium \cite{Libchaber}. In a very new study it has been shown that the dynamics of DNA is also influenced by the processes dependent on the energy derived from ATP hydrolysis \cite{Mitchison}. A series of simulation studies have also been performed to investigate the looping dynamics in active system.  Shin \textit{et al} have recently shown that in presence of self-propelled particles the loop formation in polymer become faster due to increased diffusion \cite{metzler2015njp}. In another study it has been found that looping is also faster when the polymer itself is active, having a catalytic monomer. This catalytic monomer generates a concentration gradient prompting faster diffusion of the non-catalytic monomer resulting in rapid ring-closure \cite{snigdha2014}. Such studies are extremely important as loop formation in biopolymers is an essential process in protein folding, DNA replication etc.

Experimentally there have been many attempts to study the dynamics of unfolded proteins mainly involving long-range F$\ddot{o}$rster resonance energy transfer (FRET) \cite{schuler2008,schuler2012}. In this particular technique two residues of a protein are labelled with a donor and an acceptor using fluorescence probes to study the fluctuation of the distance between them (Fig. 1). This distance is temporally correlated with a characteristic decay time, referred as reconfiguration time ($\tau_{N0}$) which is determined by fitting the long time decay of the second order intensity correlation function \cite{makarov2003, makarov2010}. To the best of our knowledge no such experimental study has been performed till date which will provide insights into the reconfiguration dynamics of a chain in an active medium. In this paper we theoretically analyze the dynamics of a single chain polymer in presence of active noise. By active noise we refer to a long-ranged temporal noise where the time correlation is independent of ambient temperature. In contrary to the recent simulation studies we find such long temporally correlated noise to result in a slower reconfiguration of a polymer chain, be it flexible or semi-flexible. Even in the presence of non-local hydrodynamics interactions in addition to the active noise, reconfiguration of the chain is slower.

The paper is arranged as follows. In section $\bf{II}$ we have introduced the model for active noise, in section $\bf{III}$ the calculation methods are discussed. The results are presented in section $\bf{IV}$ and the paper is concluded in section $\bf{V}$.

 \begin{figure}
\centering
 \includegraphics[width=0.55\textwidth]{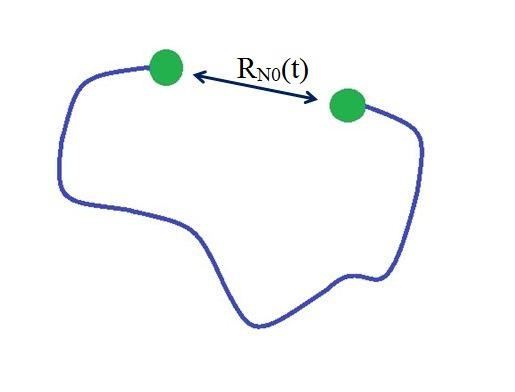}
 \caption{Schematics showing the end-to-end monomers of a protein labelled with the donor and the acceptor. The arrow depicts the distance ($R_{N0}(t)$) between the donor and acceptor monitored in the experiment.}
 \label{fig:a0}
 \end{figure}

\section{Model}

For an one-dimensional Brownian particle, moving in a harmonic trap, the dynamics in the over-damped regime is best described by the Langevin equation \cite{doibook, kawakatsubook}

\begin{equation}
\xi\frac{dx(t)}{d{t}}=-k{x(t)}+{f}(t)
\label{eq:langevin}
\end{equation}

\noindent Where, $k$ is the force constant and $f(t)$ is the Gaussian random force with first and second moments

\begin{equation}
   \left<f(t)\right>=0,
   \left<f(t^{\prime})f(t^{\prime\prime})\right>=2 \xi k_B T \delta(t^{\prime}-t^{\prime\prime})
\label{eq:random-force}
\end{equation}

\noindent Here the strength of the correlation depends on the ambient temperature $T$. Now when the system is subjected to an active noise of strength $f_A$ the equation of motion becomes \cite{ghosh2014},

\begin{equation}
\xi\frac{dx(t)}{d{t}}=-k{x(t)}+{f}(t)+{f_A}(t)
\label{eq:langevinact}
\end{equation}

\noindent Here, $f_A$ is considered to be exponentially correlated with a characteristic decay time $\tau_A$ and Gaussian distribution with moments

\begin{equation}
\left<f_A(t)\right>=0,\left\langle f_A(t^{\prime})f_A(t^{\prime\prime}) \right\rangle)=Ce^{-\frac{|t^{\prime}-t^{\prime\prime}|}{\tau_A}}
\label{eq:active-force}
\end{equation}

\noindent Importantly $C$, the strength of the active noise is independent of $T$ and can be related to probability of active force ($P(f_A)$) and the force $f_A$ acting on the particle as f $C \sim P(f_A)f_A^2$. Being independent of ambient temperature $T$ the active noise drives the system off the equilibrium and only in the infinite time limit a stationary state can be realized. In between, the system remains in a non-equilibrium state. Such a choice of noise correlation comes from earlier simulation studies on red-blood cell membrane fluctuations, where the force $f_A(t)$ originates from the non-equilibrium fluctuations of the motor proteins \cite{Shokef}.

The position correlation function $\left\langle x(t)x(0)\right \rangle$  with the active noise is analytically trackable and has the following expression. Readers are referred to the appendix for the detailed derivation.

\begin{equation}
\phi(t)=\left\langle x(t)x(0) \right\rangle=\frac{k_BT}{k}e^{-\frac{t}{\tau}}+\frac{C}{\xi^2}e^{-\frac{t}{\tau}}\left(\frac{1}{\frac{1}{\tau}\left(\frac{1}{\tau}+\frac{1}{\tau_A}\right)}+\frac{e^{\frac{t}{\tau}-\frac{t}{\tau_A}}-1}{\left(\frac{1}{\tau^2}-\frac{1}{\tau_A^2} \right)} \right)
\label{eq:corrpart}
\end{equation}

\noindent In the absence of active noise, $C \rightarrow 0$, the above expression is reduced to

$$\lim_{C\to0}\phi(t)=\frac{k_BT}{k}e^{-\frac{t}{\tau}}$$

\noindent This is the time correlation function for an over-damped Brownian particle in harmonic potential in the presence of only thermal noise or the Ornstein-Uhlenbeck process and $\tau=\frac{\xi}{k}$ is the corresponding relaxation time \cite{sokolovbook}. It is obvious from Eq. (\ref{eq:corrpart}) that although the correlation function is translationally invariant, it is not single exponential. However, this can be approximated as a single exponential with an effective relaxation time $\tau_{eff}$,

\begin{equation}
\phi_{eff}(t)=\frac{k_BT_{0}}{k}e^{-\frac{t}{\tau_{eff}}}
\end{equation}

\noindent with,

\begin{equation}
\begin{aligned}
\tau_{eff} & = \int\limits_{0}^{\infty} dt  \frac{\phi(t)}{\phi(0)} \\
& = \tau \left[\frac{\left(k_BT\xi^2+C\tau\tau_Ak\right)\left(\tau+\tau_A\right)}{C\tau^2\tau_Ak+k_BT\xi^2\left(\tau+\tau_A\right)}\right]
\end{aligned}
\end{equation}\\

\noindent $\tau_{eff}$ is bound from the above and below with $\tau$ and $\tau+\tau_A$. In the limit, $T \rightarrow \infty$, $\tau_{eff}=\tau$, in other extreme $T \rightarrow 0$, $\tau_{eff}=\tau+\tau_A$. Similarly, as $C \rightarrow \infty$, the correlation decay also become slower with $\tau_{eff}=\tau+\tau_A$ and in the absence of noise when $C \rightarrow 0$, $\tau_{eff}=\tau$. Other than $\tau_{eff}$ another parameter is $T_{0}$ that defines the effective correlation function $\phi_{eff}(t)$. $T_0$ is related to the ambient temperature as follows,   $k_BT_0=k_B\left(T+\frac{Ck\tau}{\xi^2\left(\frac{1}{\tau}+\frac{1}{\tau_A}\right)}\right)$. \\

\noindent Thus, $k_BT_0$ defines a renormalized thermal energy, but only in the limit $t \rightarrow \infty$. This directly follows from the mean square displacement (MSD) of the particle.

\begin{equation}
\left\langle\left(x(t)-x(0)\right)^2\right\rangle=2\left(\phi(0)-\phi(t)\right) \\    \\ =   
\frac{2k_BT}{k}\left(1-e^{-\frac{t}{\tau}}\right)+\frac{2C}{\xi^2\frac{1}{\tau}\left(\frac{1}{\tau}+\frac{1}{\tau_A}\right)}\left(1-e^{-\frac{t}{\tau}}\right)-\frac{2C\left(e^{-\frac{t}{\tau_A}}-e^{-\frac{t}{\tau}}\right)}{\xi^2\left(\frac{1}{\tau^2}-\frac{1}{\tau_A^2}\right)}
\label{eq:msdpart}
\end{equation}

\noindent This MSD grows with time $t$ and saturates as expected due to confinement. However, the initial growth of MSD is faster in presence of active noise ($C\neq0, \tau_A\neq0$). A detailed derivation of MSD is presented in the appendix.

 This model can be further extended to describe a many-body system such as a flexible or a semi-flexible polymer as long as the chain is assumed to have a Gaussian distribution.

\subsection{Rouse chain}

Rouse model is the simplest yet widely used model to describe a polymer with Gaussian statistics devoid of any hydrodynamics and excluded volume interaction. The equation of motion of $n^{th}$ monomer is given by  \cite{doibook, kawakatsubook}

\begin{equation}
\xi\frac{\partial{R_n}(t)}{\partial{t}}=k\frac{\partial^2{R_{n}(t)}}{\partial{n^2}}+{f}(n,t)
\label{eq:rouse-model}
\end{equation}

\noindent Where, ${R_n}(t)$ is the position of the $n^{th}$ monomer at time $t$ and $n$ can vary from $0$ to $N$ for a polymer with $(N+1)$ monomers. The friction coefficient is denoted by $\xi$, which is proportional to solvent viscosity and $k$ is the spring constant which is related to the Kuhn length, $b$ as $k=\frac{3k_BT}{b^2}$ with ${f}(n,t)$ being the random force acting on $n^{th}$ monomer at time $t$ which denotes the collision between the monomer with its surrounding solvent molecules.

\begin{equation}
   \left<f(n,t)\right>=0,
   \left<f_{\alpha}(n,t^{\prime})f_{\beta}(m,t^{\prime\prime})\right>=2 \xi k_B T \delta_{\alpha\beta}\delta(n-m)\delta(t^{\prime}-t^{\prime\prime})
\label{eq:random-forcerouse}
\end{equation}

\noindent As a simple extension of the above model one can consider a Rouse chain in presence of active noise $f_A(t)$.

\begin{equation}
\xi\frac{\partial{R_n}(t)}{\partial{t}}=k\frac{\partial^2{R_{n}(t)}}{\partial{n^2}}+{f}(n,t)+f_A(n,t)
\label{eq:rouse-model-act}
\end{equation}

It is a very standard procedure to decouple the equation of motion of the monomers using normal modes having independent motions as follows, $R_{n}(t)={X_0} + 2 \sum\limits_{p=1}^\infty X_p(t)cos(\frac{p \pi n}{N})$ and as long as the noises $f(n,t)$ and $f_A(n,t)$ are uncorrelated Eq. (\ref{eq:rouse-model-act}) converts to

\begin{equation}
\xi_p \frac{d{X_p}(t)}{d{t}}=-k_p X_{p}(t)+f_p(t)+f_{A,p}(t)
\label{eq:rouse-mode}
\end{equation}

\noindent Where, $k_p=\frac{6k_BTp^2\pi^2}{Nb^2}$ and $\xi_p=2N\xi$. The relaxation time for the $p^{th}$ normal mode in absence of any active noise is $\tau_p=\frac{\xi_p}{k_p}=\frac{\tau_1}{p^2}$, where $\tau_1 =\frac{\xi N^2 b^2}{3 k_BT\pi^2}$ is known as Rouse time. $f_{p}(t)$ and $f_{A,p}(t)$ are
random and active forces respectively which follow

\begin{equation}
\left<f_{p\alpha}(t)\right>=0,\left<f_{p\alpha}(t^{\prime})f_{q\beta}(t^{\prime\prime})\right>=2 \xi_p k_B T \delta_{\alpha\beta}\delta_{pq}\delta(t^{\prime}-t^{\prime\prime})
\label{eq:randomforce_modes}
\end{equation}

\begin{equation}
\left<f_{A,p\alpha}(t)\right>=0,\left\langle f_{A,p\alpha}(t^{\prime})f_{A,q\beta}(t^{\prime\prime}) \right\rangle)=2NC\delta_{\alpha\beta}\delta_{pq}e^{-\frac{|t^{\prime}-t^{\prime\prime}|}{\tau_A}}
\label{eq:activeforce_modes}
\end{equation}

\noindent The above equation (Eq. (\ref{eq:rouse-mode})) is structurally the same as that of Eq. (\ref{eq:langevinact}), the only difference being it is for the $p^{th}$ mode of a chain. It is obvious that each mode of the chain behaves as an over-damped Brownian particle in the presence of the active noise trapped in a harmonic well.

\subsection{Zimm chain}

When pre-averaged hydrodynamic interactions are considered under $\theta$ condition, the normal modes of the polymer behave very similarly as that of a Rouse chain and have the same structure \cite{doibook}

\begin{equation}
\xi_p^Z \frac{d{X_p}(t)}{d{t}}=-k_p^Z X_{p}(t)+f_p(t)+f_{A,p}(t)
\label{eq:zimm-mode}
\end{equation}

\noindent with $\xi_{p}^Z=\xi\sqrt{\frac{\pi N p}{3}}$ where $k_p^Z=k_p$ and $\tau_p^Z=\frac{\xi_p^Z}{k_p^Z}$ and $\tau_1^Z=\frac{\xi N^{3/2}b^2}{6\sqrt{3}\pi^{3/2}k_BT}$ \cite{chakrabartiphysica1}.

\subsection{Wormlike chain}

The semi-flexible polymer is modeled as Kratky-Porod wormlike chain which is unstretchable and includes the effect of bending energy  \cite{doibook, Liverpool2003}. The  equation of motion for a semi-flexible chain without incorporating the effects from hydrodynamic interactions is given by

\begin{equation}
\xi\frac{\partial{R_n}(t)}{\partial{t}}=k\frac{\partial^2{R_{n}(t)}}{\partial{n^2}}-\kappa\frac{\partial^4{R_{n}(t)}}{\partial{n^4}}+{f}(n,t)+f_A(n,t)
\label{eq:rousesemi-model}
\end{equation}

\noindent In normal mode description semi-flexible chain is similar to a flexible chain except $k_p$ which has a fourth order dependence on the mode number $p$ unlike flexible chain.

\begin{equation}
\xi_p^S \frac{d{X_p}(t)}{d{t}}=-k_p^S X_{p}(t)+f_p(t)+f_{A,p}(t)
\label{eq:semi-mode}
\end{equation}

\noindent Where, $k_p^S=\frac{6k_BTp^2\pi^2}{Nb^2}+\frac{2\kappa p^4\pi^4}{N^3b^3}$ and $\kappa$, bending rigidity is related to the persistence length $l_p$ of the polymer as follows $\kappa =k_BTl_p$. However, $\xi_{p}^S=\xi_{p}=2N\xi$ and $\tau_p^S=\frac{\xi_p^S}{k_p^S }$.

\section{Calculation methods}

The time-correlation function for the normal modes has a very general structure for the flexible as well as the semi-flexible chain and it remains the same even when the hydrodynamic interactions are incorporated. The form of the expression is very similar to that of a single over-damped Brownian particle moving in a harmonic well in the presence of active noise

\begin{equation}
\left\langle X_{p\alpha}(t)X_{q\beta}(0)\right\rangle=\frac{k_BT}{k_p}\delta_{\alpha\beta}\delta_{pq}e^{-\frac{t}{\tau_p}}+\frac{2 N C}{\xi_p^2}\delta_{\alpha\beta}\delta_{pq}e^{-\frac{t}{\tau_p}}\left(\frac{1}{\frac{1}{\tau_p}\left(\frac{1}{\tau_p}+\frac{1}{\tau_A}\right)}+\frac{e^{\frac{t}{\tau_p}-\frac{t}{\tau_A}}-1}{\left(\frac{1}{\tau_p^2}-\frac{1}{\tau_A^2} \right)} \right)
\label{eq:corrgen}
\end{equation}

\noindent To find the exact expression for the flexible, semi-flexible or the Zimm chain one just need to select the exact forms of $k_p$, $\xi_p$ and $\tau_p$ for a Rouse chain,  $k_p^Z$, $\xi_p^Z$, $\tau_p^Z$ for a Zimm chain and $k_p^S$, $\xi_p^S$, $\tau_p^S$ for a semi-flexible chain.

\noindent The time correlation function for the vector ($R_{N0}$) connecting the $N^{th}$ and the $0^{th}$ monomer can easily be calculated from the above expression, which is the summation over the correlation functions of all normal modes describing the polymer.

\begin{equation}
{\phi_{N0}}(t)=\left<{R}_{N0}(t).{R}_{N0}(0)\right>
=16 \sum\limits_{p=odd}^\infty 3\left\langle X_{p}(t)X_{p}(0)\right\rangle
\label{eq:normcorr}
\end{equation}

The reconfiguration time ($\tau_{N0}$) corresponding to the fluctuation of the distance between the end-to-end monomers is theoretically calculated by taking a time integration of the corresponding normalized correlation function ($\Phi_{N0}(t)$) \cite{chakrabarti2014, chakrabarti2015}

\begin{equation}
{\tau}_{N0}=\int\limits_{0}^\infty dt {\Phi}_{N0}(t)
\label{eq:recon}
\end{equation}

\noindent Where, ${\Phi}_{N0}(t)=\frac{\phi(t)}{\phi(0)}$

Similarly, the expression of the MSD of the vector ($R_{N0}$) can also be derived from the MSD of the normal modes \cite{toan,chakrabarti2014}. This is again similar to that of a single particle. In a recent study, Ghosh \textit{et al.} \cite{ghosh2014} have demonstrated how MSD of a semi-flexible chain grows in presence of such active noise. Higher the strength of the active noise, faster the growth.

\begin{equation}
\begin{aligned}
\left\langle\left(R_{N0}(t)-R_{N0}(0)\right)^2\right\rangle= 2\left(\phi_{N0}(0)-\phi_{N0}(t)\right) \\ \\ =16\sum_{p=odd}^{\infty} 3 \Bigg( \frac{2k_BT}{k_p}\left(1-e^{-\frac{t}{\tau_p}}\right)+& \frac{2C}{\xi_p^2\frac{1}{\tau_p}\left(\frac{1}{\tau_p}+\frac{1}{\tau_A}\right)}\left(1-e^{-\frac{t}{\tau_p}}\right)\\
&-\frac{2C\left(e^{-\frac{t}{\tau_A}}-e^{-\frac{t}{\tau_p}}\right)}{\xi_p^2\left(\frac{1}{\tau_p^2}-\frac{1}{\tau_A^2}\right)} \Bigg)
\end{aligned}
\label{eq:MSD}
\end{equation}

\section{Results and discussions}

\begin{figure}
\centering
 \includegraphics[width=0.8\textwidth]{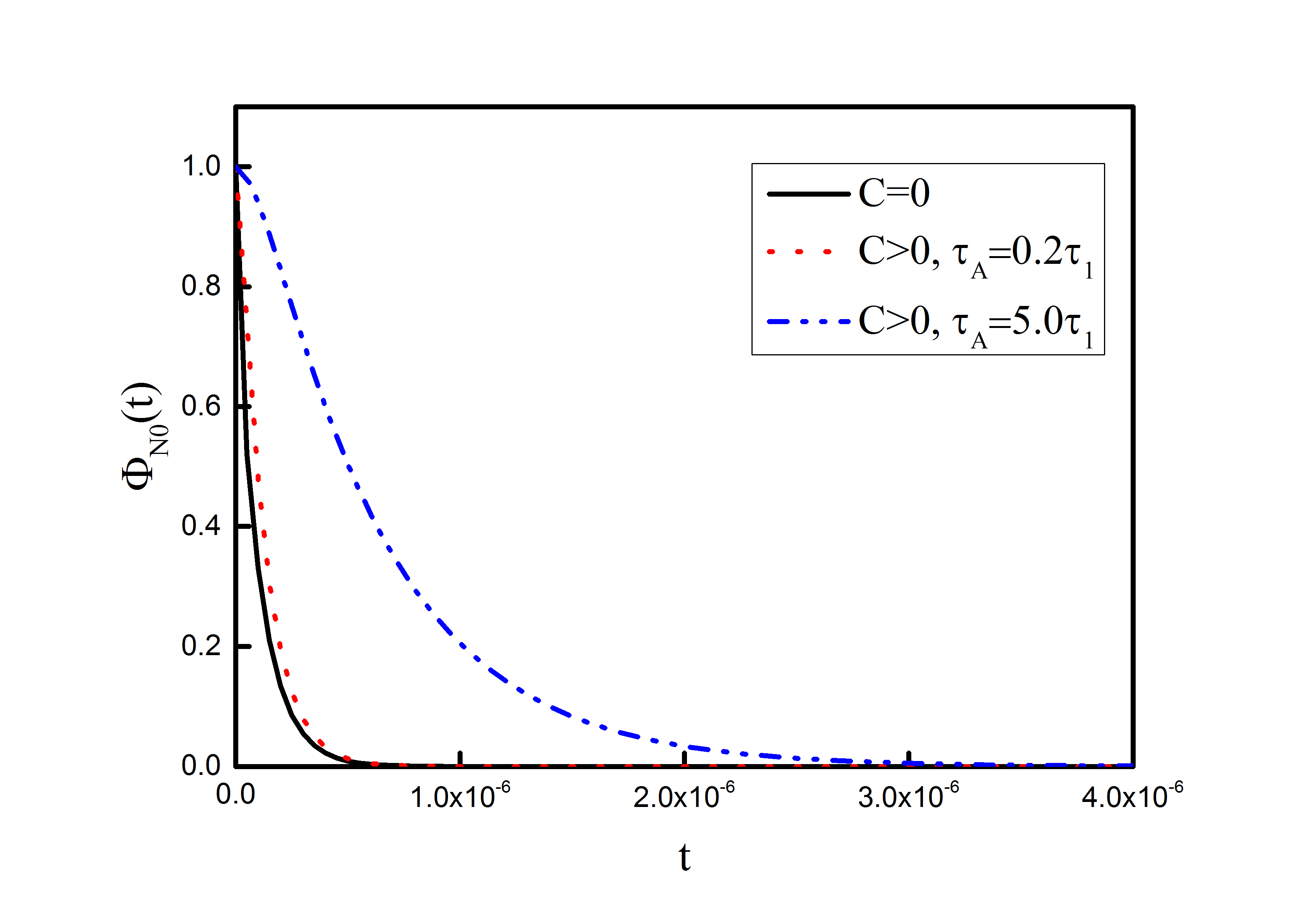}
 \caption{Plot of $\Phi_{N0}(t)$ vs $t$ for Rouse chain}
 \label{fig:a}
 \end{figure}

\begin{figure}
\centering
 \includegraphics[width=0.8\textwidth]{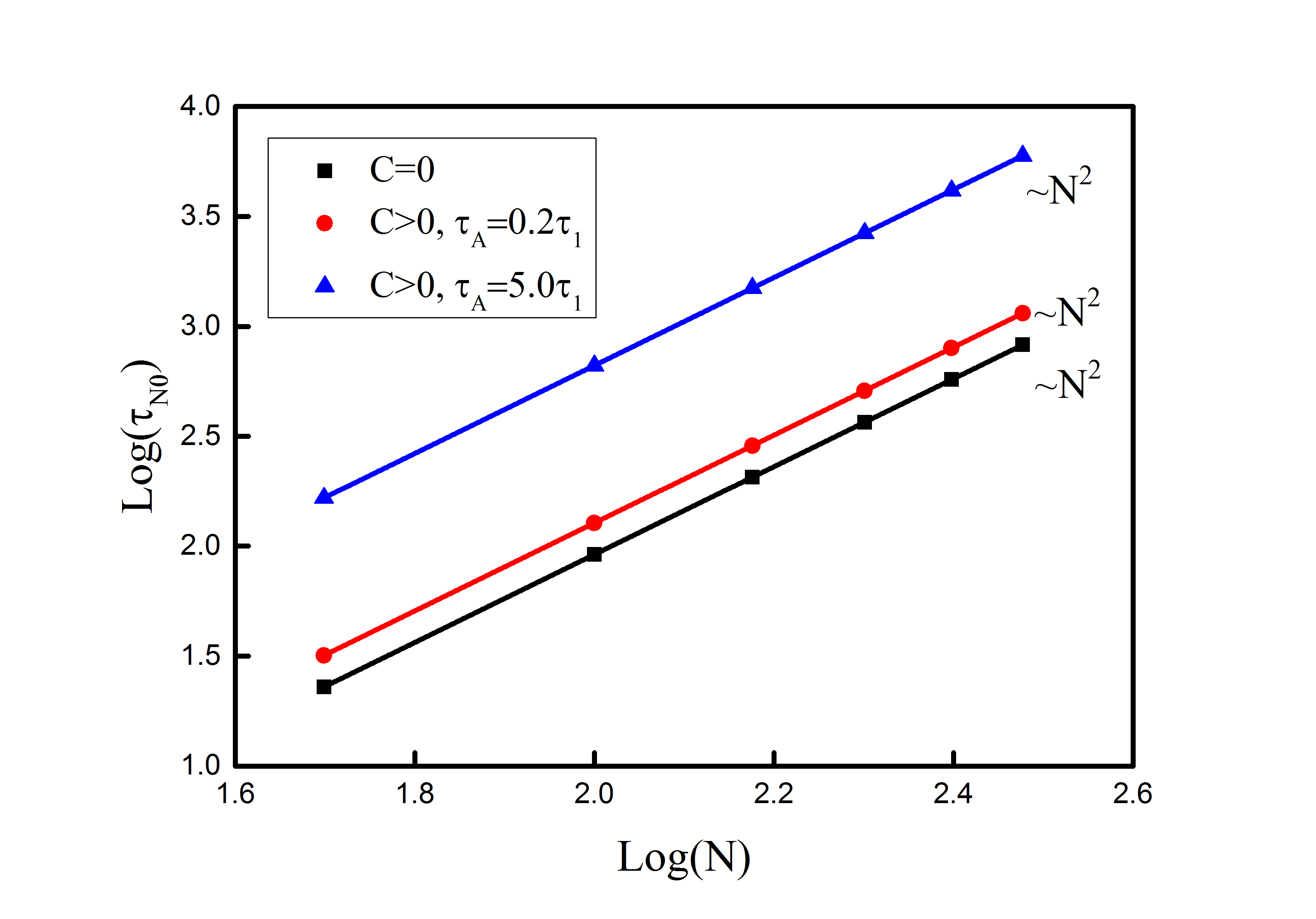}
 \caption{Log-log plot of reconfiguration time ($\tau_{N0}$) vs Chain length ($N$) for Rouse chain }
 \label{fig:b}
 \end{figure}

In Fig. (\ref{fig:a}) we show the normalized correlation function for Rouse chain in presence and absence of active noise which is calculated using the generalized expression given in Eq. (\ref{eq:corrgen}). The parameters are chosen in consistence with the real values such as $N = 100$, $b=3.8\times10^{-10} m$, $k_B = 1.38\times10^{-23} JK^{-1}$, $T = 300 K$ and $\xi = 9.42\times10^{-12} kgs^{-1}$ which is in agreement with the viscosity of water. As mentioned earlier $C$ is the strength of the active noise and $C=\frac{P(f_A) f_A^2}{b}$. It has been experimentally observed that in biological systems motor proteins like myosin, kinesin exert force in the $\sim5-10 pN$ range \cite{Cell_Biology_book}. For our calculations we have considered $f_a = 10\times10^{-12} N$ and $P(f_A)=1$. For a fixed value of $C$ we have chosen two different values of $\tau_A$, such as $0.2\tau_1$ and  $5.0\tau_1$ which are in the same order of magnitude of $\tau_1$. From the plot it can be seen that correlation decay is always slower in presence of an active noise even when the characteristic decay time of the active noise $\tau_A$ is very small and as $\tau_A$ increases the decay of $\Phi_{N0}$ becomes even slower. However, this correlation loss has very weak dependence on the strength of the active noise $C$. Changing the strength practically brings no difference in the correlation function. The log-log plot of reconfiguration time against chain length ($N$) is shown in Fig (\ref{fig:b}) where the reconfiguration time is calculated using Eq. (\ref{eq:recon}) and as expected, reconfiguration time increases as the temporal correlation loss of the active noise becomes slower. For, a $100$ monomer chain $\tau_{N0}$ increases $\sim 1.4$ times in the presence of active noise when $\tau_A=0.2\tau_1$, whereas it becomes $\sim 7$ times higher when the decay time of active noise $\tau_A=5\tau_1$. But, surprisingly the chain length dependence remains unchanged even in the presence of active noise. In all three cases $\nu=2$ where, $\tau_{N0}\sim N^{\nu}$. It is well known that the reconfiguration time is a summation of the relaxation times of each mode i.e. $\tau_{N0}=16 \sum\limits_{p=odd}^\infty 3\left\langle X_{p}(t)X_{p}(0)\right\rangle$, which has the analytically exact expression

\begin{equation}
\tau_{N0}=16 \sum\limits_{p=odd}^{\infty}3{\tau}_p \left[\frac{\left(k_BT\xi_p^2+C{\tau}_p\tau_Ak_p\right)\left({\tau}_p+\tau_A\right)}{C{\tau}_p^2\tau_Ak_p+k_BT\xi_p^2\left({\tau}_p+\tau_A\right)}\right]
\label{eq:recon_gen}
\end{equation}

\noindent Where, the $N$ dependence comes through ${\tau}_p$, $\xi_p$ and $k_p$. A careful analysis of the preceding expression shows that if the active noise strength $C$ is very small i.e. $C \rightarrow 0$, the above expression reduces to $\tau_{N0}\simeq 16 \sum\limits_{p=odd}^{\infty}3{\tau}_p$ and since, $\tau_p \sim N^2$, the dependence of the reconfiguration time is also identical. Now what happens if $C$ becomes very large i.e. $C \rightarrow \infty$, $\tau_{N0}\simeq 16 \sum\limits_{p=odd}^{\infty}3{\tau}_p+\tilde{\tau}$, where $\tilde{\tau}$ is a constant. Even in this case the $N$ dependence comes only from $\tau_p$ and $\tau_{N0}\sim N^2$. In between these two extreme cases the active noise cause very small change in the $N$ dependence of the reconfiguration time which is reflected Fig (\ref{fig:b}).

\begin{figure}
\centering
 \includegraphics[width=0.8\textwidth]{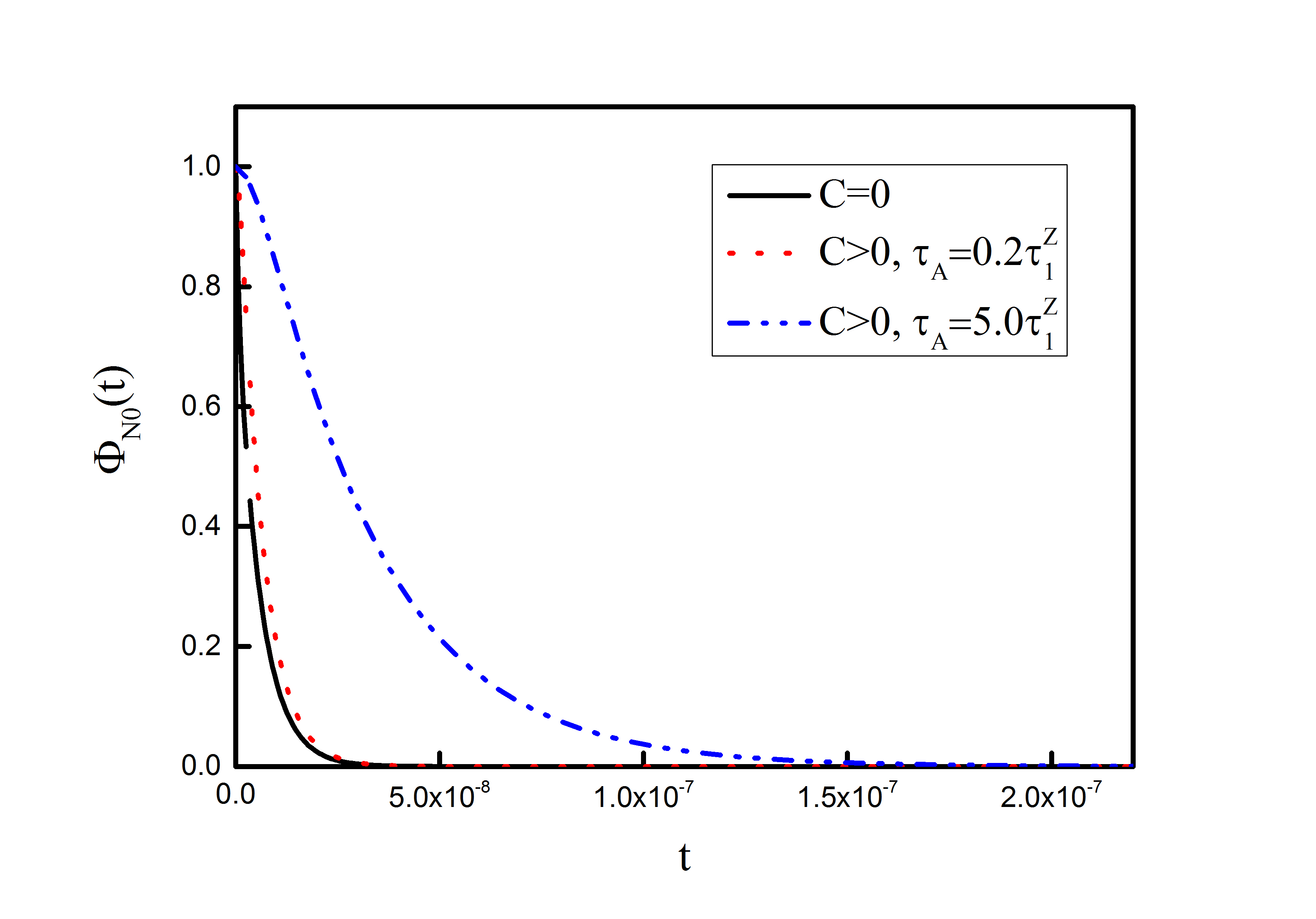}
 \caption{Plot of $\Phi_{N0}(t)$ vs $t$ for Zimm chain}
 \label{fig:c}
 \end{figure}

\begin{figure}
\centering
 \includegraphics[width=0.8\textwidth]{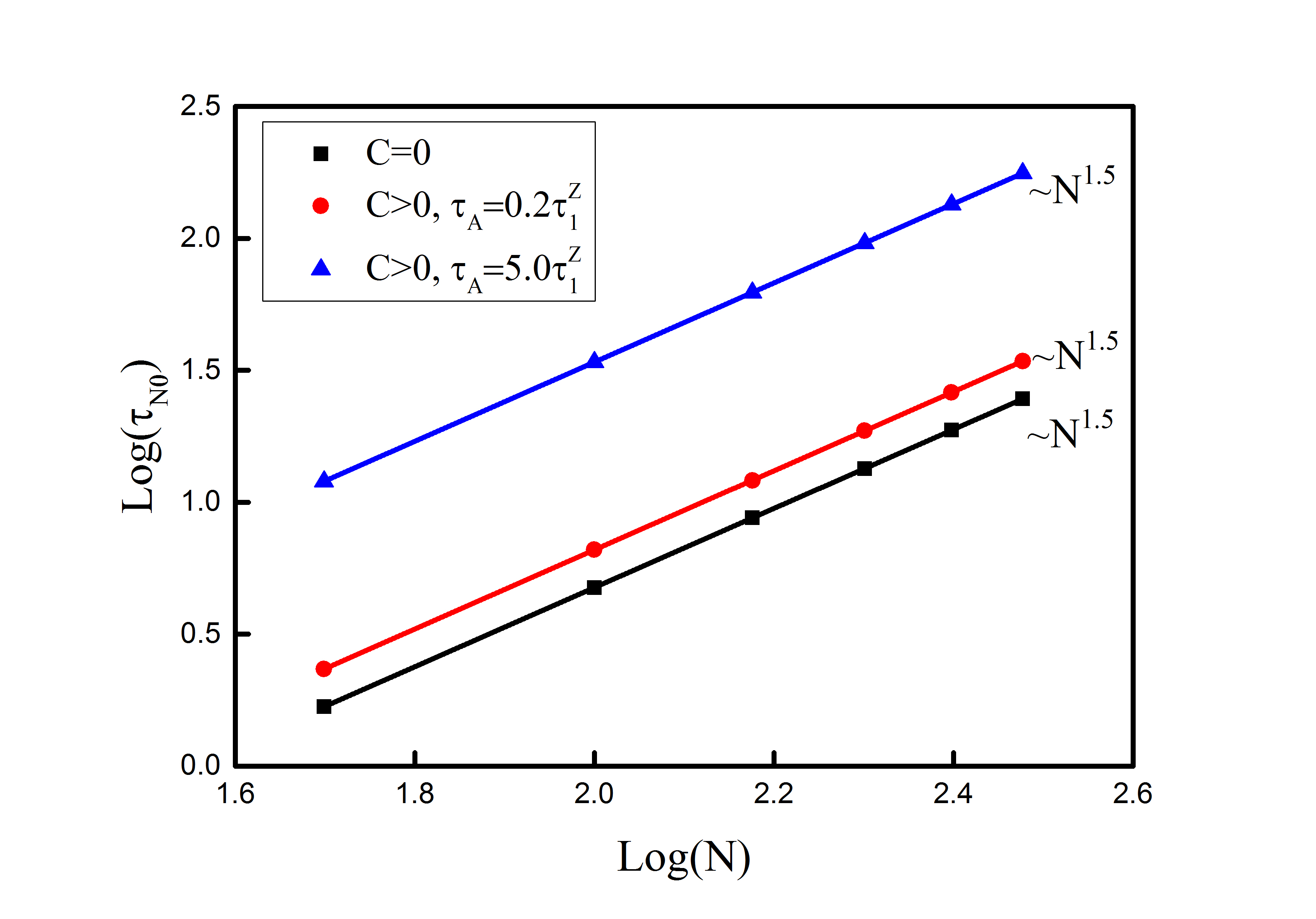}
 \caption{Log-log plot of reconfiguration time ($\tau_{N0}$) vs Chain length ($N$) for Zimm chain }
 \label{fig:d}
 \end{figure}

The same set of calculations have been performed for a flexible polymer including the pre-averaged hydrodynamic interaction under $\theta$ condition. The plot of normalized time-correlation function against time is shown in Fig. (\ref{fig:c}) which shows a similar trend as that of Rouse chain, i.e. the correlation loss becomes slower whenever active noise is introduced to the system. Next, the chain length dependence of the reconfiguration time is determined for the Zimm chain from Fig. (\ref{fig:d}), and it is found to be $\sim N^{1.5}$ which is in agreement to the previous work done by Chakrabarti \cite{chakrabartiphysica1}. In this case also the chain-length dependence of reconfiguration time does not differ in the presence of active noise.

\begin{figure}
\centering
 \includegraphics[width=0.8\textwidth]{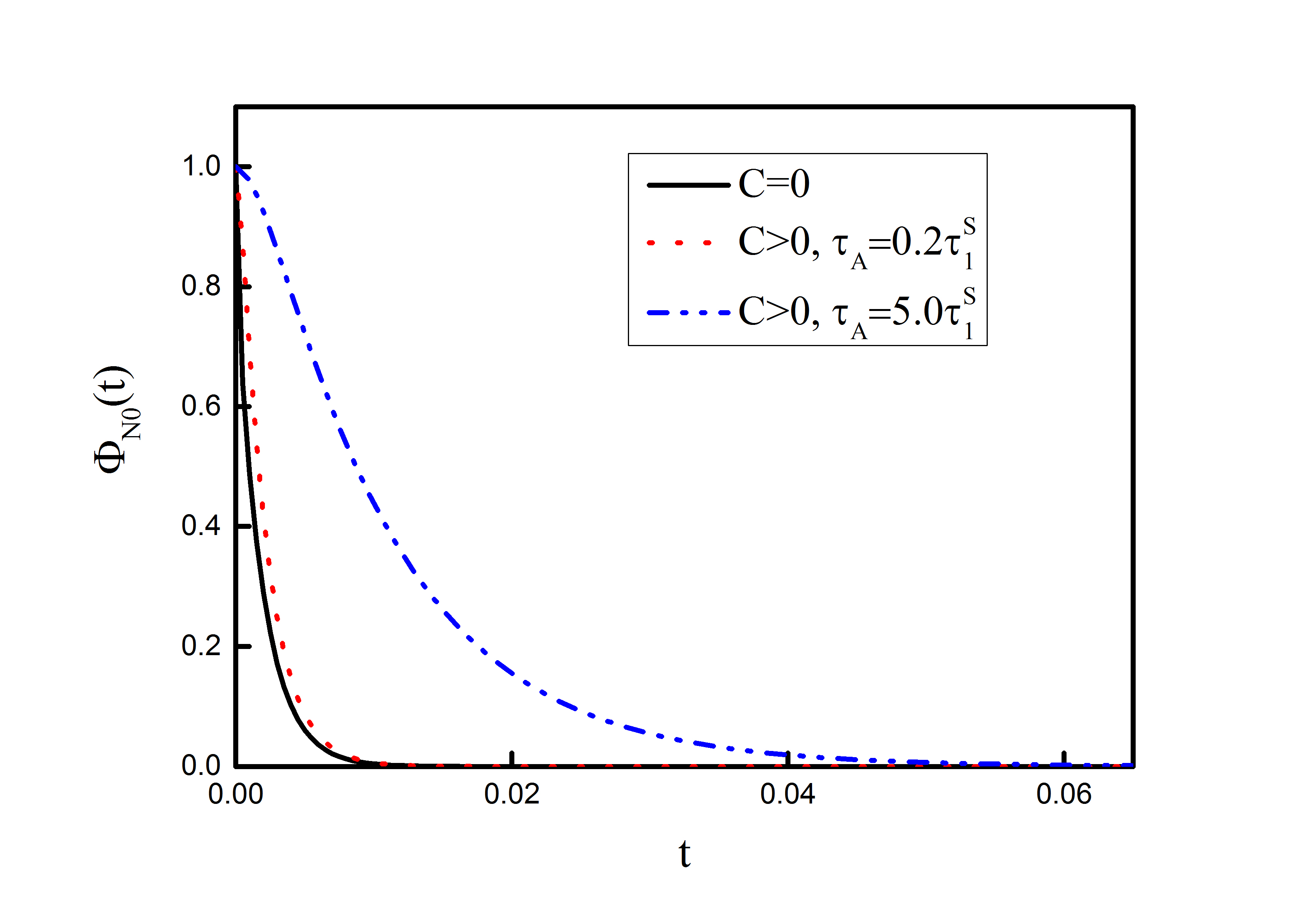}
 \caption{Plot of $\Phi_{N0}(t)$ vs $t$ for semi-flexible chain}
 \label{fig:e}
 \end{figure}

Fig. (\ref{fig:e}) shows the normalized time correlation function of the end-to-end vector for a semi-flexible chain. Here also the behavior of the correlation loss in presence of active noise is identical to Rouse and Zimm chain. The Kuhn length for semi-flexible has been considered to be $b = 50\times10^{-9} m$. This is roughly the Kuhn length of DNA which has a series of different Kuhn length depending upon the solvent condition \cite{Manning}. The persistence length of the semi-flexible chain has been considered to be half of the Kuhn length during the calculations. When chain length dependence of semi-flexible chain is determined it is found to be $\sim N^{2}$. This might seem surprising since $\tau_p^S$ has a fourth order dependence on $N$ in addition to the usual second order dependence, for which the dependence on chain length should be higher than that of Rouse model. However, if we take a look at the expression for $\tau_p^S$, since the value of $\kappa$ considered in the calculations is very small the contribution from $N^4$ is negligibly small and that is the reason even for semi-flexible chain the $\tau_{N0}\sim N^2$.

 \begin{figure}
\centering
 \includegraphics[width=0.8\textwidth]{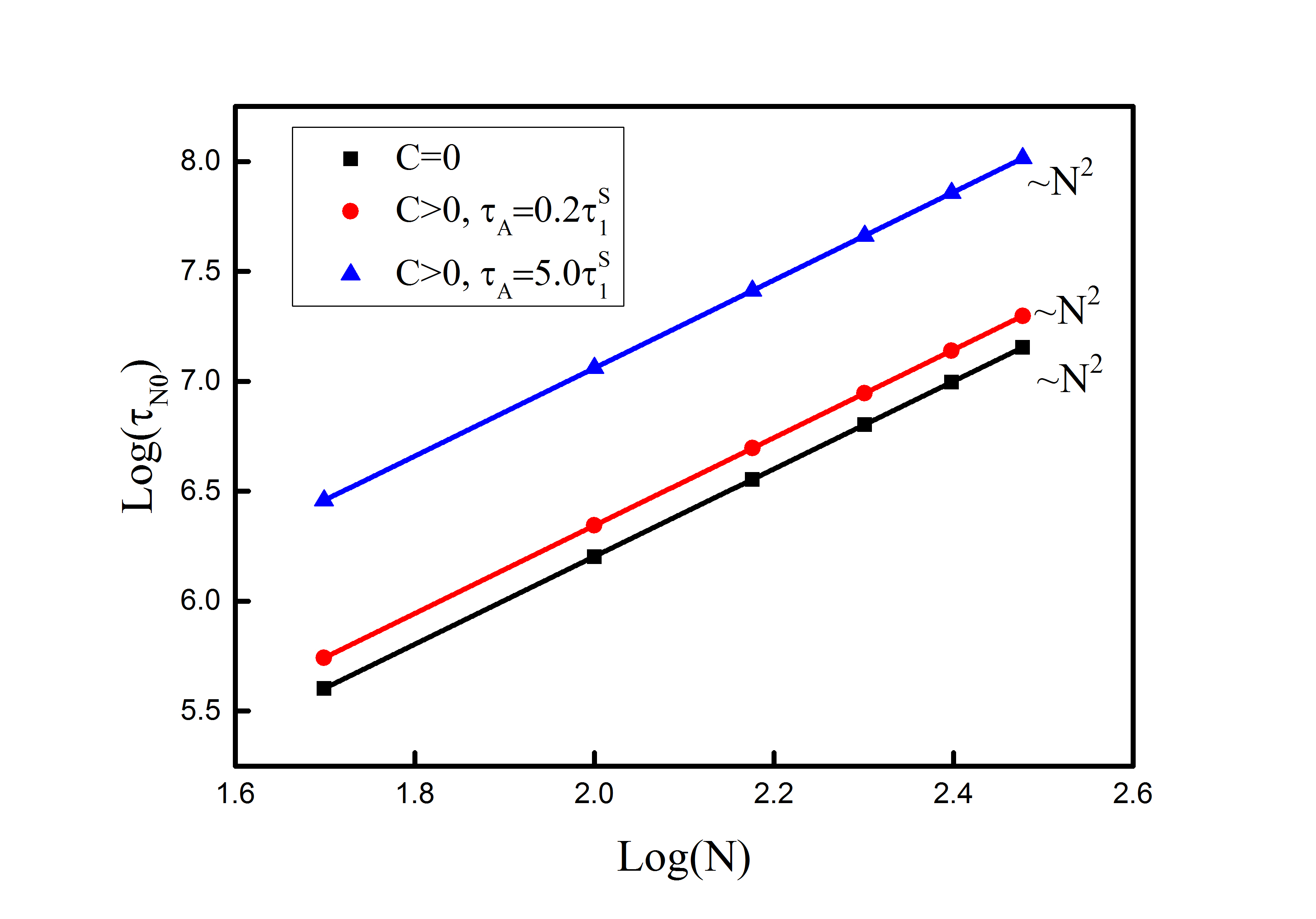}
 \caption{Log-log plot of reconfiguration time ($\tau_{N0}$) vs Chain length ($N$) for semi-flexible chain }
 \label{fig:f}
 \end{figure}

\section{Conclusion}

In this work we have looked into the effect of active noise in chain reconfiguration for a flexible polymer where the active noise is modelled with a long temporally correlated non-equilibrium force. It can be clearly seen that in the presence of active noise the chain reconfiguration becomes slower. However slowing down of the reconfiguration dynamics seems to be controlled by the correlation time $\tau_A$ rather than the strength of the correlation. Thus in a typical FRET like experiment, measurement of reconfiguration time of a protein should show slowing down in an environment with active noise. Similar behaviors are also observed when pre-averaged hydrodynamic interaction is considered. For a worm-like semi-flexible chain the trend remains the same. However, the dependence on chain-length of the reconfiguration time does not change in the presence of active noise. Keeping long story short, our study suggests that not always presence of active noise can guarantee faster reconfiguration of a polymer chain. In an environment, where a long temporal noise acts on the chain, FRET type measurement would show the chain to retain the correlation for longer time than in absence of such noise.

\section{Appendix}
\subsection{Correlation function}

\noindent The equation of motion for a single over-damped Brownian particle trapped in harmonic potential in presence of active noise

%$\xi\frac{d{{x}(t)}}{d{t}}=-k{x}(t)+{f}(t)+f_A(t)$

$\frac{d{{x}(t)}}{d{t}}+\frac{k}{\xi}{x}(t)=\frac{1}{\xi}\left({f}(t)+f_A(t)\right)$\\

\noindent Multiplying the integrating factor $e^{\frac{k}{\xi}t}$ on both sides we get,

$\left(\frac{d{{x}(t)}}{d{t}}+\frac{k}{\xi}{x}(t)\right)e^{\frac{k}{\xi}t}=\frac{1}{\xi}\left({f}(t)e^{\frac{k}{\xi}t}+f_A(t)e^{\frac{k}{\xi}t}\right)$\\

\noindent Integrating boths side from $-\infty$ to $t$ (which means we assume the system to start evolving at infinite past).

$\int\limits_{-\infty}^t\frac{d}{dt}\left(x(t^{\prime})e^{\frac{k}{\xi}t^{\prime}}\right)=\frac{1}{\xi}\int\limits_{-\infty}^t\left({f}(t^{\prime})e^{\frac{k}{\xi}t^{\prime}}+f_A(t^{\prime})e^{\frac{k}{\xi}t^{\prime}}\right)$

or, $x(t)=\frac{e^{-\frac{t}{\tau}}}{\xi}\int\limits_{-\infty}^t dt^{\prime}\left({f}(t^{\prime})e^{\frac{t^{\prime}}{\tau}}+f_A(t^{\prime})e^{\frac{t^{\prime}}{\tau}}\right)$ as, $\frac{\xi}{k}=\tau$

\noindent Since the thermal and the active noise are uncorrelated, they come separately as a summation in the correlation function and the position correlation function for the thermal noise has standard solution which is not shown here,

\begin{equation*}
\begin{split}
\left\langle x(t)x(0) \right\rangle & = \frac{e^{-\frac{t}{\tau}}}{\xi^2}\int\limits_{-\infty}^t dt^{\prime}\int\limits_{-\infty}^0dt^{\prime\prime}\left(\left\langle f(t^{\prime})f(t^{\prime\prime}) \right\rangle + \left\langle f_A(t^{\prime})f_A(t^{\prime\prime}) \right\rangle\right) e^{\left(\frac{t^{\prime}+t^{\prime\prime}}{\tau}\right)}\\
& = \frac{k_BT}{k}e^{-\frac{t}{\tau}}+\frac{C}{\xi^2}e^{-\frac{t}{\tau}}\int\limits_{-\infty}^t dt^{\prime}\int\limits_{-\infty}^0dt^{\prime\prime}e^{\frac{t^{\prime}+t^{\prime\prime}}{\tau}} e^{-\frac{|t^{\prime}-t^{\prime\prime}|}{\tau_A}}
\end{split}
\end{equation*}

%or, $\left\langle x(t)x(0) \right\rangle=\frac{k_BT}{k}e^{-\frac{t}{\tau}}+\frac{C}{\xi^2}e^{-\frac{t}{\tau}}\int\limits_{-\infty}^t dt^{\prime}\int\limits_{-\infty}^0dt^{\prime\prime}e^{\frac{t^{\prime}+t^{\prime\prime}}{\tau}} e^{-\frac{|t^{\prime}-t^{\prime\prime}|}{\tau_A}}$\\

\noindent The position correlation function for The time correlation function of the active noise involves a modulus of time, therefore the integration is split in two parts. One where $t^{\prime}>t^{\prime\prime}$ and another considering $t^{\prime}<t^{\prime\prime}$.

\begin{figure}
\centering
 \includegraphics[width=0.8\textwidth]{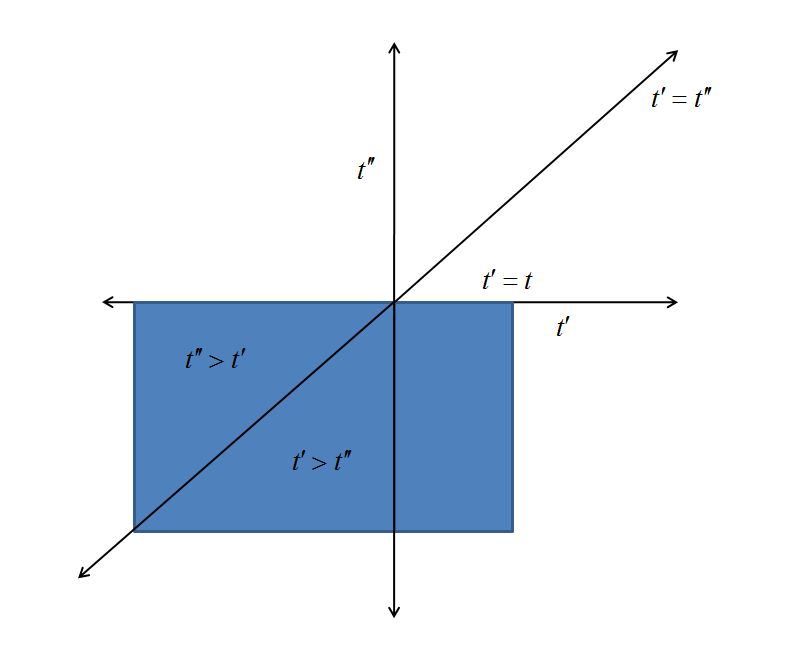}
 \caption{The shaded region represents the range of time that has to be integrated for active noise to calculate the time correlation function.}
 \label{fig:g}
 \end{figure}

\begin{equation*}
\begin{aligned}
\left\langle x(t)x(0) \right\rangle=&\frac{k_BT}{k}e^{-\frac{t}{\tau}}+\frac{C}{\xi^2}e^{-\frac{t}{\tau}} \Bigg( \int\limits_{t^{\prime\prime}=-\infty}^0 dt^{\prime\prime}\int\limits_{t^{\prime}=t^{\prime\prime}}^{0}dt^{\prime}e^{\left( \frac{t^{\prime}}{\tau}-\frac{t^{\prime}}{\tau_A} \right)} e^{\left( \frac{t^{\prime\prime}}{\tau}+\frac{t^{\prime\prime}}{\tau_A} \right)}+ \\
&\int\limits_{t^{\prime}=-\infty}^0 dt^{\prime}\int\limits_{t^{\prime\prime}=t^{\prime}}^{0}dt^{\prime\prime}e^{\left( \frac{t^{\prime\prime}}{\tau}-\frac{t^{\prime\prime}}{\tau_A} \right)} e^{\left( \frac{t^{\prime}}{\tau}+\frac{t^{\prime}}{\tau_A} \right)}+\int\limits_{t^{\prime\prime}=-\infty}^0 dt^{\prime\prime}\int\limits_{t^{\prime}=0}^{t}dt^{\prime}e^{\left( \frac{t^{\prime}}{\tau}-\frac{t^{\prime}}{\tau_A} \right)} e^{\left( \frac{t^{\prime\prime}}{\tau}+\frac{t^{\prime\prime}}{\tau_A} \right)}\Bigg) \\
&=\frac{k_BT}{k}e^{-\frac{t}{\tau}}+\frac{C}{\xi^2}e^{-\frac{t}{\tau}}\Bigg(2\int\limits_{t^{\prime\prime}=-\infty}^0 dt^{\prime\prime}\int\limits_{t^{\prime}=t^{\prime\prime}}^{0}dt^{\prime}e^{\left( \frac{t^{\prime}}{\tau}-\frac{t^{\prime}}{\tau_A} \right)} e^{\left( \frac{t^{\prime\prime}}{\tau}+\frac{t^{\prime\prime}}{\tau_A} \right)}+\\
& \int\limits_{t^{\prime\prime}=-\infty}^0 dt^{\prime\prime}\int\limits_{t^{\prime}=0}^{t}dt^{\prime}e^{\left( \frac{t^{\prime}}{\tau}-\frac{t^{\prime}}{\tau_A} \right)} e^{\left( \frac{t^{\prime\prime}}{\tau}+\frac{t^{\prime\prime}}{\tau_A} \right)}\Bigg)\\
&=\frac{k_BT}{k}e^{-\frac{t}{\tau}}+\frac{C}{\xi^2}e^{-\frac{t}{\tau}}\left(\frac{2}{\left(\frac{1}{\tau}-\frac{1}{\tau_A} \right)}\int\limits_{t^{\prime\prime}=-\infty}^0dt^{\prime\prime}\left(e^{\frac{2t^{\prime\prime}}{\tau}}-e^{\frac{t^{\prime\prime}}{\tau}+\frac{t^{\prime\prime}}{\tau_A}}\right)+\frac{1}{\left(\frac{1}{\tau}+\frac{1}{\tau_A} \right)}\frac{e^{\frac{t}{\tau}-\frac{t}{\tau_A}}-1}{\left(\frac{1}{\tau}-\frac{1}{\tau_A} \right)}\right)
\end{aligned}
\end{equation*}

%\begin{dmath}\left\langle x(t)x(0) \right\rangle=\frac{k_BT}{k}e^{-\frac{t}{\tau}}+\frac{C}{\xi^2}e^{-\frac{t}{\tau}}\left(2\int\limits_{t^{\prime\prime}=-\infty}^0 dt^{\prime\prime}\int\limits_{t^{\prime}=t^{\prime\prime}}^{0}dt^{\prime}e^{\left( \frac{t^{\prime}}{\tau}-\frac{t^{\prime}}{\tau_A} \right)} e^{\left( \frac{t^{\prime\prime}}{\tau}+\frac{t^{\prime\prime}}{\tau_A} \right)}+\int\limits_{t^{\prime\prime}=-\infty}^0 dt^{\prime\prime}\int\limits_{t^{\prime}=0}^{t}dt^{\prime}e^{\left( \frac{t^{\prime}}{\tau}-\frac{t^{\prime}}{\tau_A} \right)} e^{\left( \frac{t^{\prime\prime}}{\tau}+\frac{t^{\prime\prime}}{\tau_A} \right)}\right)\end{dmath}

%\begin{dmath}\left\langle x(t)x(0) \right\rangle=\frac{k_BT}{k}e^{-\frac{t}{\tau}}+\frac{C}{\xi^2}e^{-\frac{t}{\tau}}\left(\frac{2}{\left(\frac{1}{\tau}-\frac{1}{\tau_A} \right)}\int\limits_{t^{\prime\prime}=-\infty}^0dt^{\prime\prime}\left(e^{\frac{2t^{\prime\prime}}{\tau}}-e^{\frac{t^{\prime\prime}}{\tau}+\frac{t^{\prime\prime}}{\tau_A}}\right)+\frac{1}{\left(\frac{1}{\tau}+\frac{1}{\tau_A} \right)}\frac{e^{\frac{t}{\tau}-\frac{t}{\tau_A}}-1}{\left(\frac{1}{\tau}-\frac{1}{\tau_A} \right)}\right)\end{dmath}

\begin{equation}\left\langle x(t)x(0) \right\rangle=\frac{k_BT}{k}e^{-\frac{t}{\tau}}+\frac{C}{\xi^2}e^{-\frac{t}{\tau}}\left(\frac{1}{\frac{1}{\tau}\left(\frac{1}{\tau}+\frac{1}{\tau_A}\right)}+\frac{e^{\frac{t}{\tau}-\frac{t}{\tau_A}}-1}{\left(\frac{1}{\tau^2}-\frac{1}{\tau_A^2} \right)} \right)
\label{eq:corrpart1}
\end{equation}

\subsection{MSD}

The mean-square displacement(MSD) of single particle in one-dimension is as follows,
$\left\langle\left(x(t)-x(0)\right)^2\right\rangle$

$x(t)=\frac{e^{-\frac{t}{\tau}}}{\xi}\int\limits_{-\infty}^t dt^{\prime}\left({f}(t^{\prime})e^{\frac{t^{\prime}}{\tau}}+f_A(t^{\prime})e^{\frac{t^{\prime}}{\tau}}\right)$

$x(t)-x(0)=\frac{e^{-\frac{t}{\tau}}}{\xi}\int\limits_{-\infty}^t dt^{\prime}\left({f}(t^{\prime})e^{\frac{t^{\prime}}{\tau}}+f_A(t^{\prime})e^{\frac{t^{\prime}}{\tau}}\right)-\frac{1}{\xi}\int\limits_{-\infty}^0 dt^{\prime}\left({f}(t^{\prime})e^{\frac{t^{\prime}}{\tau}}+f_A(t^{\prime})e^{\frac{t^{\prime}}{\tau}}\right)$

\begin{equation*}
\begin{aligned}
\left\langle\left(x(t)-x(0)\right)^2\right\rangle=&\frac{e^{-\frac{2t}{\tau}}}{\xi^2}\int\limits_{-\infty}^t dt^{\prime}\int\limits_{-\infty}^t dt^{\prime\prime}\left(\left\langle f(t^{\prime})f(t^{\prime\prime}) \right\rangle + \left\langle f_A(t^{\prime})f_A(t^{\prime\prime}) \right\rangle\right) e^{\left(\frac{t^{\prime}+t^{\prime\prime}}{\tau}\right)}+\\
&\frac{1}{\xi^2}\int\limits_{-\infty}^0 dt^{\prime}\int\limits_{-\infty}^0 dt^{\prime\prime}\left(\left\langle f(t^{\prime})f(t^{\prime\prime}) \right\rangle + \left\langle f_A(t^{\prime})f_A(t^{\prime\prime}) \right\rangle\right) e^{\left(\frac{t^{\prime}+t^{\prime\prime}}{\tau}\right)}-\\
&2\frac{e^{-\frac{t}{\tau}}}{\xi^2}\int\limits_{-\infty}^t dt^{\prime}\int\limits_{-\infty}^0 dt^{\prime\prime}\left(\left\langle f(t^{\prime})f(t^{\prime\prime}) \right\rangle + \left\langle f_A(t^{\prime})f_A(t^{\prime\prime}) \right\rangle\right) e^{\left(\frac{t^{\prime}+t^{\prime\prime}}{\tau}\right)}
\end{aligned}
\end{equation*}

\noindent Again the MSD for an over-damped Brownian particle in presence of harmonic potential is well-known which is not shown here in detail,

\begin{equation*}
\begin{aligned}
\frac{e^{-\frac{2t}{\tau}}}{\xi^2} \int\limits_{-\infty}^t dt^{\prime}\int\limits_{-\infty}^t  dt^{\prime\prime}\big(\left\langle f(t^{\prime})f(t^{\prime\prime}) \right\rangle & + \left\langle f_A(t^{\prime})f_A(t^{\prime\prime}) \right\rangle \big) e^{\left(\frac{t^{\prime}+t^{\prime\prime}}{\tau}\right)}=\\
&\frac{k_BT}{k}+\frac{C}{\xi^2}\Bigg(e^{-\frac{2t}{\tau}}\left(\frac{1}{\frac{1}{\tau}\left(\frac{1}{\tau}+\frac{1}{\tau_A}\right)}+2\frac{e^{\frac{t}{\tau}-\frac{t}{\tau_A}}-1}{\left(\frac{1}{\tau^2}-\frac{1}{\tau_A^2}\right)} \right)+\\
&\frac{1}{\frac{1}{\tau}\left(\frac{1}{\tau}+\frac{1}{\tau_A} \right)}\left(1-e^{-\frac{2t}{\tau}}\right)-\frac{2e^{-\frac{t}{\tau}}}{\left(\frac{1}{\tau^2}-\frac{1}{\tau_A^2} \right)}\left(e^{-\frac{t}{\tau_A}}-e^{-\frac{t}{\tau}}\right)\Bigg)
\end{aligned}
\end{equation*}

\begin{equation*}
\begin{aligned}
\frac{1}{\xi^2}\int\limits_{-\infty}^0 dt^{\prime}\int\limits_{-\infty}^0 dt^{\prime\prime}\left(\left\langle f(t^{\prime})f(t^{\prime\prime}) \right\rangle + \left\langle f_A(t^{\prime})f_A(t^{\prime\prime}) \right\rangle\right) e^{\left(\frac{t^{\prime}+t^{\prime\prime}}{\tau}\right)}=\\
&\frac{k_BT}{k}+\frac{C}{\xi^2}\left(\frac{1}{\frac{1}{\tau}\left(\frac{1}{\tau}+\frac{1}{\tau_A}\right)}\right)
\end{aligned}
\end{equation*}

\begin{equation*}
\begin{aligned}
2\frac{e^{-\frac{t}{\tau}}}{\xi^2} \int\limits_{-\infty}^t dt^{\prime}\int\limits_{-\infty}^0 dt^{\prime\prime}\big(\left\langle f(t^{\prime})f(t^{\prime\prime}) \right\rangle & + \left\langle f_A(t^{\prime})f_A(t^{\prime\prime}) \right\rangle\big) e^{\left(\frac{t^{\prime}+t^{\prime\prime}}{\tau}\right)}=\\
&\frac{2k_BT}{k}e^{-\frac{t}{\tau}}+\frac{2C}{\xi^2}e^{-\frac{t}{\tau}}\left(\frac{1}{\frac{1}{\tau}\left(\frac{1}{\tau}+\frac{1}{\tau_A}\right)}+\frac{e^{\frac{t}{\tau}-\frac{t}{\tau_A}}-1}{\left(\frac{1}{\tau^2}-\frac{1}{\tau_A^2} \right)} \right)
\end{aligned}
\end{equation*}

\begin{equation}
\left\langle\left(x(t)-x(0)\right)^2\right\rangle=\frac{2k_BT}{k}\left(1-e^{-\frac{t}{\tau}}\right)+\frac{2C}{\xi^2\frac{1}{\tau}\left(\frac{1}{\tau}+\frac{1}{\tau_A}\right)}\left(1-e^{-\frac{t}{\tau}}\right)-\frac{2C\left(e^{-\frac{t}{\tau_A}}-e^{-\frac{t}{\tau}}\right)}{\xi^2\left(\frac{1}{\tau^2}-\frac{1}{\tau_A^2}\right)}
\label{eq:msdpart1}
\end{equation}

%\begin{bibliography}
%\bibliographystyle{apsrev}
%\usepackage[square,numbers,sort&compress]{natbib}
%\bibliographystyle{apsrev}
%\usepackage{doi}
%\usepackage{hyperref}

%\bibliography{ringclosurenew}
%\end{bibliography}

\end{document}